\documentclass[showpacs,pre]{revtex4}
\usepackage{graphicx}
\begin{document}
\title{Ratio control in a cascade model of cell differentiation}
\author{Hidetsugu Sakaguchi}
\affiliation{Department of Applied Science for Electronics and Materials,
Interdisciplinary Graduate School of Engineering Sciences, 
Kyushu University, Kasuga, Fukuoka 816-8580, Japan}
\begin{abstract}
We propose a kind of reaction-diffusion equations for cell differentiation, which exhibits the Turing instability. If the diffusivity of some variables is set to be infinity, we get coupled competitive reaction-diffusion equations with a global feedback term.  The size ratio of each cell type is controlled by a system parameter in the model. Finally, we extend the model to a cascade model of cell differentiation. A hierarchical spatial structure appears as a result of the cell differentiation. The size ratio of each cell type is also controlled by the system parameter. 
\end{abstract}
\maketitle
\section{Introduction}
Reaction-diffusion systems are important models for various nonlinear processes in nonequilibrium physics and biological systems~\cite{rf:1}. 
Turing first proposed a reaction-diffusion model to explain pattern formation in developmental biology~\cite{rf:2}. Various pattern formation has been studied in theoretical models~\cite{rf:3,rf:4}. 

There is a characteristic wavelength in the Turing pattern. Kondo and Asai observed that the number of stripes in the skin pattern of a tropical fish increases  to keep the characteristic wavelength constant, when the fish grows with time~\cite{rf:5}. In the skin pattern, the periodic pattern is constructed of pigment cells with different colors. The different types of pigment cells are mutually competitive.

On the other hand, there are several observations that various types of cells are differentiated during the developmental process, but the number ratio of the different cell types does not change very much as the body size increases.  A typical example is the cellular slime mold Dictyostelium discoideum. The amoebic state changes to the fruiting body via the migrating slug state, if the breeding condition becomes worse. Prespore cells and prestalk cells are differentiated in the process. The number ratio is kept almost constant even if the body size is changed~\cite{rf:6}.  The ratio control and the pattern formation are performed in two steps. In an early stage, the ratio of the two-type cells is regulated by  DIF (Differentiation Inducing Factor), and later prestalk cells move into a tip region within the slug through a chemotaxis by the cAMP~\cite{rf:7,rf:8}.  The two cell types, i.e., the prespore and the prestalk  are competitive in this system. Another example is the differentiation of blood cells from the stem cells in the bone marrow. The number ratio of the red blood cells, the white blood cells, and the platelets is kept to be roughly constant.

 Another typical rule of the cell differentiation is a cascade control of the differentiation process by complicated gene networks. In the segmentation process of the Drosophila, the differentiation proceeds from a large scale to a small scale in a hierarchical manner.  A cascade network of genes and proteins such as bicoid protein, gap genes and pair-rule genes are identified in detail~\cite{rf:9}. Several competitive relations between two genes are known also in this process. For example, the gene engrailed (en) for the posterior compartmental specification and the gene wingless (wg) for the anterior compartment are mutually competitive. Several theoretical models of the hierarchical gene network were proposed for the segmentation process~\cite{rf:10,rf:11}.

The analyses of the specific gene network and the pattern formation for each system are important in the developmental biology. However, in this paper, we consider a very simple cascade model of competitive reaction-diffusion equations and propose a mechanism of the ratio control from a view point of the nonlinear dynamics. 

\begin{figure}[t]
\begin{center}
\includegraphics[width=9cm]{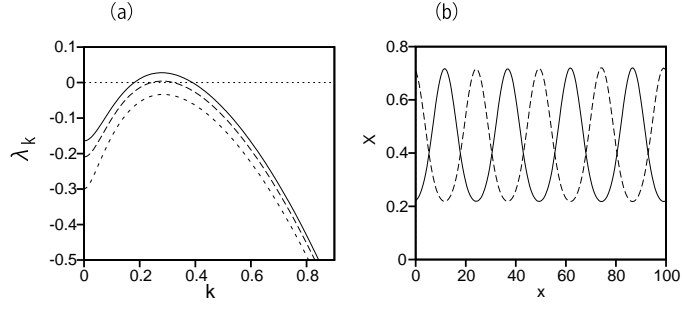}
\end{center}
\caption{(a) Eigenvalue $\lambda_k$ for Eq.~(2) at  $a=0.4,c=2,d=1,D_X=1,D_Y=20$ and $b=5$ (solid curve), $b=4$ (dashed curve), and $b=3$ (dotted curve).
(b) Stationary profiles of $X_1$ (solid curve) and $X_2$ (dashed curve) at $a=0.4,b=5,c=2,d=1$ and $D_X=1$. 
}
\label{f1}
\end{figure}
\begin{figure}[t]
\begin{center}
\includegraphics[width=14cm]{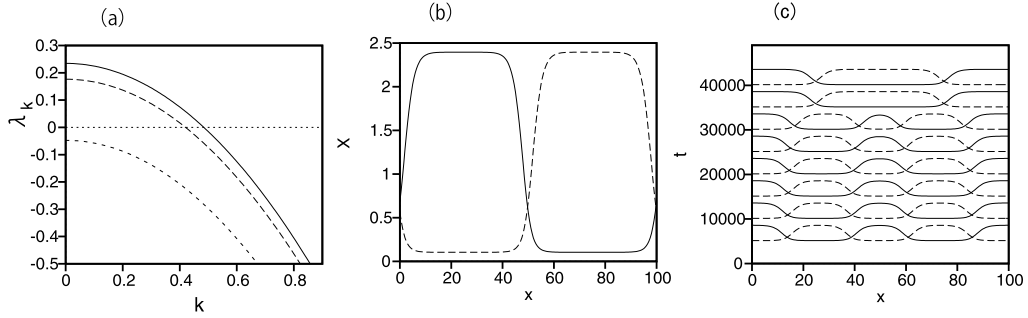}
\end{center}
\caption{(a) Eigenvalue $\lambda_k$ for Eq.~(4) at  $a=0.4,c=2,d=1,D_X=1$ and $b=5$ (solid curve), $b=3$ (dashed curve), and $b=1$ (dotted curve).
(b) Stationary profiles of $X_1$ (solid curve) and $X_2$ (dashed curve) at $a=0.4,b=5,c=2,d=1,D_X=1$. (c) Time evolution of $X_1$ (solid curve) and $X_2$ (dashed curve) from  $X_1=0.4+0.04\cos(2\pi x/L)+0.1\cos(4\pi x/L)$ and $X_2=0.4-0.1\cos(4\pi x/L)$.
}
\label{f2}
\end{figure}
\begin{figure}[t]
\begin{center}
\includegraphics[width=9cm]{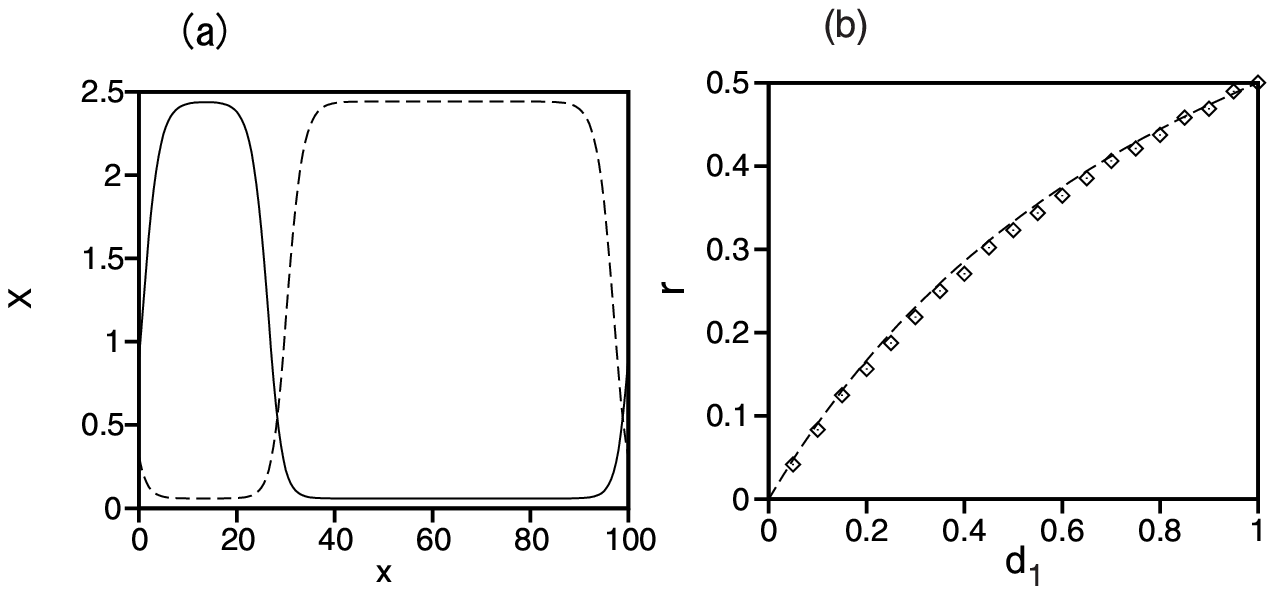}
\end{center}
\caption{(a) Stationary profiles of $X_1$ (solid curve) and $X_2$ (dashed curve) at $a=0.4,b=5,c=2,d_1=0.4,d_2=1$ and $D_X=1$ for Eq.~(6). (b) Size ratio of the $X_1$-dominant region as a function of $d_1$ for $a=0.4,b=5,c=2,d_1=0.4,d_2=1$ and $D_X=1$. The rhombi denote numerical results and the dashed curve is $d_1/(d_1+d_2)$.
}
\label{f3}
\end{figure}
\section{Ratio-control in a competitive reaction-diffusion system}
If two cell-types are mutually competitive, only one type of cells are locally  activated by the lateral inhibition. The two cell-types are bistable, and a homogeneous state of one-type cell will appear. However, if some long-range enhancement of the competitive type is included in the system, two types of cells will appear in a large system. Meinhardt and Gierer proposed a model of the mutual induction of locally exclusive states, and generated a stripe-like pattern~\cite{rf:12}.  We study first a simplified version of their model.
The model is written as coupled reaction-diffusion equations:
\begin{eqnarray}
\frac{\partial X_1}{\partial t}&=&\frac{cX_1+dY_2}{1+aX_1+bX_2}-X_1+D_X\nabla^2X_1,\nonumber\\ 
\frac{\partial X_2}{\partial t}&=&\frac{cX_2+dY_1}{1+aX_2+bX_1}-X_2+D_X\nabla^2X_2,\nonumber\\ 
\frac{\partial Y_1}{\partial t}&=&X_1-Y_1+D_Y\nabla^2Y_1,\nonumber\\
\frac{\partial Y_2}{\partial t}&=&X_2-Y_2+D_Y\nabla^2Y_2,
\end{eqnarray}
where $a,b,c,d$ are positive parameters, $X_1$ and $X_2$ obey the Hill type equation, $Y_1$ and $Y_2$ are produced respectively from $X_1$ and $X_2$. The Hill type equation is often used in the biochemistry and the gene regulatory network~\cite{rf:13,rf:14}.  The enhancement and the suppression of the reaction are expressed in the numerator and the denominator in the first term of on the right-hand side of the Hill equation. That is, both $X_1$ and $X_2$ acts as activators for themselves by the terms of $cX_1$ or $cX_2$, and acts as inhibitors for the other by the term of $bX_2$ and $bX_1$. The two components $X_1$ and $X_2$ compete with each other when $b$ is sufficiently large. The terms $aX_1$ and $aX_2$ in the denominators express the effect of the saturation. 
We assume that $X_1$ and $X_2$ correspond to proteins (morphogens) determining competitive cell types such as the pigment cells of different colors in the skin pattern, or the prestalk and the prespore in the slime mould. The term of $dY_2$ and $dY_1$ implies that $Y_2$ and $Y_1$ are activators respectively for $X_1$ and $X_2$. We assume that  $D_Y\gg D_X$. It means that the morphogens $Y_1$ and $Y_2$ diffuse rapidly and work as long-range activators respectively for $X_2$ and $X_1$.  

There is a uniform solution $X_1=X_2=Y_1=Y_2=X_0=(c+d-1)/(a+b)$ to Eq.~(1).  
It is expected that $X_2$ is depressed when $X_1$ increases locally and vice versa, because $X_1$ and $X_2$ are mutually competitive. 
If the perturbation of the form: $X_1=X_0+\delta x_1\cos(kx),  X_2=X_0-\delta x_1\cos(kx), Y_1=X_0+\delta y_1\cos(kx), Y_2=X_0-\delta y_1\cos(kx)$ are assumed owing to the competitive relation, the deviations $\delta x_1$ and $\delta y_1$ satisfy
\begin{eqnarray}
\frac{d\delta x_1}{dt}&=&\frac{c\delta x_1-d\delta y_1}{1+(a+b)X_0}+\frac{(c+d)(b-a)X_0\delta x_1}{\{1+(a+b)X_0\}^2}-(1+D_Xk^2)\delta x_1,\nonumber\\
\frac{d\delta y_1}{dt}&=&\delta x_1-(1+D_Yk^2)\delta y_1.
\end{eqnarray}
The eigenvalue $\lambda_k$ can be calculated from Eq.~(2). Figure 1(a) displays  $\lambda_k$ for $b=5$ (solid curve), $b=4$ (dashed curve) and $b=3$ (dotted curve) at $a=0.4,c=2,d=1,D_X=1$ and $D_Y=20$. (The parameter values $a,b,c,d,D_X$ and $D_y$ are not realistic biological ones, but we study the nonlinear system as just a model system in this paper.)
The Turing type instability is expected for a finite wavenumber. We have performed a one-dimensional numerical simulation. The system size $L=100$, and the periodic boundary conditions are imposed. Figure 1(b) shows stationary profiles of $X_1(x)$ (solid curve) and $X_2(x)$ (dashed curve) at $a=0.4,b=5,c=2,d=1,D_X=1$ and $D_Y=20$. A spatially periodic pattern appears, in which the $X_1$-dominant region and $X_2$-dominant region reciprocate spatially. This is a Turing pattern in the competitive reaction-diffusion systems. Note that the $Y$ component with large diffusivity acts as activators for the competitor, and it facilitates the competitor in spatially distant regions. It is different from the usual Turing pattern in usual activator-inhibitor systems, where the inhibitor with large diffusivity inhibits the activator.  The mechansim of the pattern formation by the long-range activation of the competitor is essentially the same as the model proposed by Meinhardt and Gierer almost 30 years ago~\cite{rf:12}. 
Kondo and Asai performed a numerical simulation of the usual type of activator-inhibitor system to explain the skin pattern of the tropical fish~\cite{rf:5}, but our model might be more plausible for the pattern formation, because the pigment cells are mutually competitive. 

Next, we generalize the model equation (1) to a model equation in which the whole system is separated into two regions of $X_1$-dominant region and $X_2$-dominant region. If $D_Y$ is infinitely large and a steady state is obtained for the $Y$ component,  Eq.~(1) is reduced to   
\begin{eqnarray}
\frac{\partial X_1}{\partial t}&=&\frac{cX_1+d\langle X_2\rangle}{1+aX_1+bX_2}-X_1+D_X\nabla^2X_1,\nonumber\\ 
\frac{\partial X_2}{\partial t}&=&\frac{cX_2+d\langle X_1\rangle}{1+aX_2+bX_1}-X_2+D_X\nabla^2X_2,
\end{eqnarray}
where $\langle X\rangle$ is the spatial average of $X$, i.e., $\langle X\rangle=(1/L)\int_0^LXdx$ in a one-dimensional system and $\langle X\rangle=(1/L^2)\int_0^L\int_0^LXdxdy$ in a two-dimensional system. It is because $Y$ becomes uniform owing to the infinitely large diffusivity of $Y$ in Eq.~(1). 
This is our original model. For this model equation, the small deviation $\delta x_1$ from the uniform solution satisfies
\begin{equation}
\frac{d\delta x_1}{dt}=\frac{c\delta x_1}{1+(a+b)X_0}+\frac{(c+d)(b-a)X_0\delta x_1}{\{1+(a+b)X_0\}^2}-(1+D_Xk^2)\delta x_1,
\end{equation}
for $k\ne 0$, because $\delta y_1$ is assumed to be zero owing to the uniformity of $Y$.  On the other hand, for $k=0$, 
\begin{equation}
\frac{d\delta x_1}{dt}=\frac{(c-d)\delta x_1}{1+(a+b)X_0}+\frac{(c+d)(b-a)X_0\delta x_1}{\{1+(a+b)X_0\}^2}-\delta x_1.
\end{equation}
Fugure 2(a) displays the eigenvalues $\lambda_k$ for $b=5$ (solid curve), $b=3$ (dashed curve) and $b=1$ (dotted curve) at $a=0.4,c=2,d=1,D_X=1$. 
The eigenvalue $\lambda_k$ at $k=0$ is negative for these parameter values, therefore, the Fourier mode with $k=0$ is stable. The uniform solution is unstable for $b=5$ and $b=3$.  The eigenvalue takes the largest value for the smallest wave number $k=2\pi/L$. Figure 2(b) shows a stationary solution for $b=5$ for the one-dimensional system with periodic boundary conditions of size $L=100$. 
The initial condition was $X_1=0.5-0.1(x/L)$ and $X_2=0.4+0.1(x/L)$. 
For this initial condition, the whole space is separated into the $X_1$-dominant region at $x<L/2$ and the $X_2$-dominant region at $x>L/2$. 
There are domain walls between the two domains. However, even if the initial condition is random, the whole space is separated into the two regions with the same size, although the position of the $X_1$-dominant region becomes random. It is because $\lambda_k$ takes a maximum value at $k=2\pi/L$. 
If the initial condition is $X_1=0.4+0.04\cos(2\pi x/L)+0.1\cos(4\pi x/L)$ and $X_2=0.4-0.1\cos(4\pi x/L)$, four domains, i.e., two $X_1$-dominant regions and two $X_2$-dominant regions appear initially, however, there is an attractive interaction between the domain walls and finally the two-domain structure is obtained as shown in Fig.2(c).  The time evolution is similar to the coarsening process in the one-dimensional Ginzburg-Landau (GL) equation: $\partial u/\partial t=u-u^3+\partial^2u/\partial x^2$~\cite{rf:15}. However, it is different from the GL system in that the final state is a uniform state in the GL system and the final state is a two-domain state in our system.  It is a unique point in our model that the two-domain structure of equal size appears naturally for any system size $L$.

Equation~(3) can be generalized into an asymmetric model:
\begin{eqnarray}
\frac{\partial X_1}{\partial t}&=&\frac{cX_1+d_1\langle X_2\rangle}{1+aX_1+bX_2}-X_1+D_X\nabla^2X_1,\nonumber\\ 
\frac{\partial X_2}{\partial t}&=&\frac{cX_2+d_2\langle X_1\rangle}{1+aX_2+bX_1}-X_2+D_X\nabla^2X_2,
\end{eqnarray}
where $d_1$ and $d_2$ are assumed to take different values.  
Figure 3(a) diplays stationary profiles of $X_1$ (solid curve) and $X_2$ (dashed curve) at $a=0.4,b=5,c=2,d_1=0.4,d_2=1,$ and $D_X=1$ for $L=100$. The sizes of the $X_1$-dominant region and the $X_2$-dominant region are not the same in this asymmetric model. However, the maximum and the minimum values of $X_1$ and $X_2$ and the width of the domain walls are almost the same, which are expressed as $X_{max}$ and $X_{min}$. If $d_1\langle X_2\rangle$ and $d_2\langle X_1\rangle$ are assumed to takes constant values and $d_1\langle X_2\rangle<d_2\langle X_1\rangle$, a domain wall between a $X_1$-dominant region and a $X_2$-dominant region moves as the $X_1$-dominant region shrinks, which is similar to the motion of the domain wall in the asymmetric Ginzburg-Landau equation: $\partial u/\partial t=u+\epsilon u^2-u^3+\partial^2u/\partial x^2$.  However, if the $X_1$-dominant region shrinks, $\langle X_1\rangle$ decreases and $\langle X_2\rangle$ increases in time in our system. This is due to a negative feedback effect involved in our system. 
Finally, the domain walls become stationary, when $d_1\langle X_2\rangle=d_2\langle X_1\rangle$ is satisfied. If the system size $L$ is sufficiently large, the ratio $r=l/L$ of the domain size $l$ of the $X_1$-dominated region is evaluated from the condition $d_1\langle X_2\rangle=d_2\langle X_1\rangle$  as
\begin{equation}
d_1\{(1-r)X_{max}+rX_{min}\}=d_2\{rX_{max}+(1-r)X_{min}\},
\end{equation}
because $\langle X_2\rangle\sim (1-r)X_{max}+rX_{min}$ and $\langle X_1\rangle\sim rX_{max}+(1-r)X_{min}$, if the width of domain walls is not taken into consideration. If $X_{min}\ll X_{max}$, $r$ is approximated at $r=d_1/(d_1+d_2)$. 
Thus, the size ratio is determined by the ratio of the parameters $d_1$ and $d_2$ and does not depend on the system size $L$. We studied the control of the domain size in the Ginzburg-Landau type equation in the previous work, which was the same mechanism as the present one~\cite{rf:16}. 
Figure 3(b) displays  numerically obtained size ratio $r$'s as a function of $d_1$ at $a=0.4,b=5,c=2,d_2=1$, and $D_X=1$ for $L=100$. The dashed curve denotes $r=1/(d_1+1)$. Fairly good agreement is seen. 
The ratio control can be attained by the choice of the parameters $d_1$ and $d_2$. This is a mechanism of the ratio control of our differentiation model. 
The feedback effect via the $Y$-variable with infinitely large diffusivity controls the ratio of the different cell types.  In this asymmetric model, a two-domain structure appears naturally, and the ratio of the domain size is uniquely determined by the system parameters. It might be applicable to the ratio control in the early stage of the slime mold. 

\begin{figure}[t]
\begin{center}
\includegraphics[width=14cm]{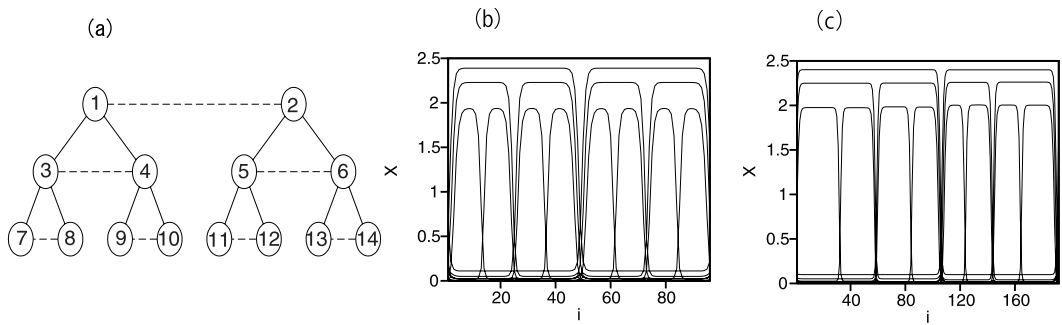}
\end{center}
\caption{(a) Hierarchical network of fourteen elements in three layers. 
The solid lines denote the active interaction and the dashed lines denote the competitive interaction. 
(b) Stationary profiles of $X_i$ ($i=1,\cdots,14$) in the cascade model similar to Eq.~(8) in a one-dimensional space. The system is composed of fourteen elements in three layers. The system size is $L=96$. The parameter values are $a=0.4,b=5,c=2,d=1,\alpha=0.4$ and $D_X=0.1$. (c) Stationary profiles of $X_i$ ($i=1,\cdots,14$) at $a=0.4,b=5,c=2,d_1=0.8,d_2=1,\alpha=0.4$ and $D_X=0.1$.
}
\label{f4}
\end{figure}
\section{A cascade model of cell differentiation}
Next, we construct a new cascade model of the competitive reaction-diffusion equations as shown in Fig.~4(a) based on Eqs.~(3) and (6), although another cascade model based on a more complicated version of Eq.~(1) was proposed by Meinhardt to study the hierarchical subdivision into gap-gene, pair-rule gene and segment polarity gene in the Drosophila~\cite{rf:10}. 
In Fig.~4(a), a system of fourteen elements in  three layers is shown.  Each element denotes a protein (morphogen) with number $i$ which determines the $i$th cell type.  The horizontal dashed lines represent the competitive interaction between two elements as in the previous model, and the solid lines from the upper layer to the lower layer represent an active interaction from a upper-level element to a lower-level element. 
A negative feedback from the lower-level to the upper level might be important, but the feedback effect is not considered in our model for the sake of simplicity.  The active and competitive interactions are represented by the Hill type equations. 
For example, a model equation for a system of six elements in two layers is written as
\begin{eqnarray}
\frac{\partial X_1}{\partial t}&=&\frac{cX_1+d_1\langle X_2\rangle}{1+aX_1+bX_2}-X_1+D_X\nabla^2X_1,\nonumber\\ 
\frac{\partial X_2}{\partial t}&=&\frac{cX_2+d_2\langle X_1\rangle}{1+aX_2+bX_1}-X_2+D_X\nabla^2X_2.\nonumber\\ 
\frac{\partial X_3}{\partial t}&=&\frac{\alpha X_1(cX_3+d_1\langle X_4\rangle)}{1+aX_3+bX_4}-X_3+D_X\nabla^2X_3,\nonumber\\ 
\frac{\partial X_4}{\partial t}&=&\frac{\alpha X_1(cX_4+d_2\langle X_3\rangle)}{1+aX_4+bX_3}-X_4+D_X\nabla^2X_4.\nonumber\\ 
\frac{\partial X_5}{\partial t}&=&\frac{\alpha X_2(cX_5+d_1\langle X_6\rangle)}{1+aX_5+bX_6}-X_5+D_X\nabla^2X_5,\nonumber\\ 
\frac{\partial X_6}{\partial t}&=&\frac{\alpha X_2(cX_6+d_2\langle X_5\rangle)}{1+aX_6+bX_5}-X_6+D_X\nabla^2X_6, 
\end{eqnarray}
where $\alpha$ is the coupling constant of the active interaction from the upper layer to the lower layer. Owing to the competitive interaction, $X_2$ is almost zero in the $X_1$-dominant region.  At the domain,  $X_5$ and $X_6$ are also almost zero because $X_5$ and $X_6$ are activated by $X_2$. On the other hand, the $X_1$-dominant region is separated into the $X_3$-dominant region and the $X_4$-dominant region because of the competitive interaction between $X_3$ and $X_4$. The size ratio of the two regions are also determined by the ratio of $d_1$ and $d_2$.   We have performed numerical simulations of a system of fourteen elements in the three layers. It is a one-dimensional system with system size $L=96$.  Figure 4(b) displays stationary profiles of $X_i$ ($i=1,\cdots,14$) for the three layer system at $a=0.4,b=5,c=2,d=1,\alpha=0.4$ and $D_X=0.1$.  
Firstly, the whole space is separated into the two domains: a $X_1$-dominant region and a $X_2$-dominant region. Next, the $X_1$-dominant region is separated into the two domains: a $X_3$-dominant region and a $X_4$-dominant region, and the $X_2$-dominant region is separated into the two domains: a $X_5$-dominant region and a $X_6$-dominant region. Similarly, the $X_3$-dominant region is separated into $X_7$-dominant and $X_8$-dominant regions, the $X_4$-dominant region is separated into $X_9$-dominant and $X_{10}$-dominant regions, the $X_5$-dominant region is separated into $X_{11}$-dominant and $X_{12}$-dominant regions, and, the $X_6$-dominant region is separated into $X_{13}$-dominant and $X_{14}$-dominant regions
As a result, a hierarchical pattern appears. The domain size decreases as $1/2,1/4$ and 1/8 as the layer is lowered.  This is a kind of binary-tree decomposition of the whole space. If the parameter $\alpha$ is assumed to take a different value  $\alpha_k$ for each layer, and is set to be $\alpha_k=(c+d/2)/\{X_{max}(c+d/2^k)\}$ for the $k$th layer, peak values of $X_i$ take almost the same value $X_{max}$ for each layer, because $\langle X_i\rangle=X_{max}/2^{k}$ in the $k$th layer. For the well tuned parameter values of $\alpha_k$, a self-similar binary decomposition of dominant regions will occur. If $d_1\ne d_2$, a  multi-fractal-like pattern appears, because  the size ratio is $r,1-r$ in the first layer, $r^2,r(1-r),(1-r)^2$ in the second layer, and $r^3,r^2(1-r),r(1-r)^2,(1-r)^3$ in the third layer.  Figure 4(b) displays a stationary profiles of $X_i$ ($i=1,\cdots, 14$)  for $d_1=0.8$ and $d_2=1$ at $a=0.4,b=5,c=2,d=1,\alpha=0.4$ and $D_X=0.1$. Inhomogeneous decomposition of dominant regions is clearly seen. 
\begin{figure}[t]
\begin{center}
\includegraphics[width=14cm]{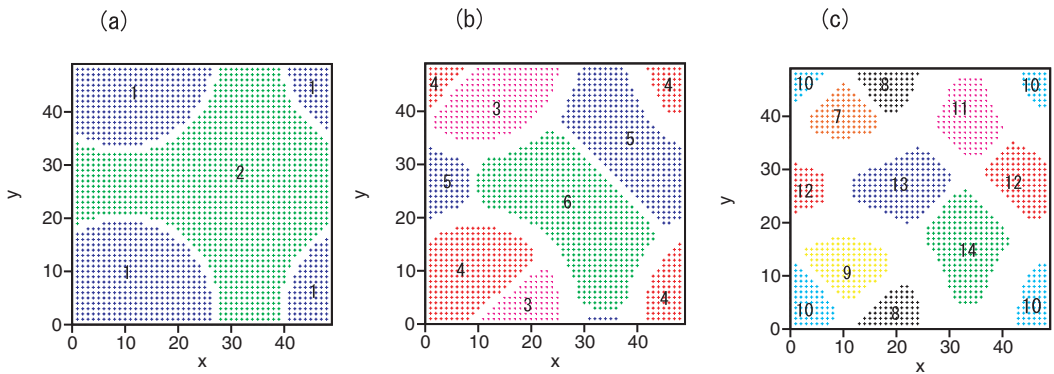}
\end{center}
\caption{Hierarchical differentiation of the fourteen elements with the three layers in a two-dimensional space. The $X_i$-dominant regions ($i=1,\cdots, 14$) are depicted with different colors in a square of size $48\times 48$.  The parameter values are $d_1=0.8$ and $d_2=1$ at $a=0.4,b=5,c=2,d_1=0.4,d_2=1,\alpha=0.4$ and $D_X=0.2$. A hierarchical structure is clearly seen. The colored regions imply that $X_i$ satisfies $X_i>1$. (a) $X_1$-dominant and $X_2$-dominant regions. (b) $X_3$, $X_4$, $X_5$, and $X_6$-dominant regions. (c) $X_7$, $X_8$, $X_9$, $X_{10}$, $X_{11}$, $X_{12}$, $X_{13}$, and $X_{14}$-dominant regions. The number $i$ indicates the cell type.
}
\label{f5}
\end{figure}
\begin{figure}[t]
\begin{center}
\includegraphics[width=5cm]{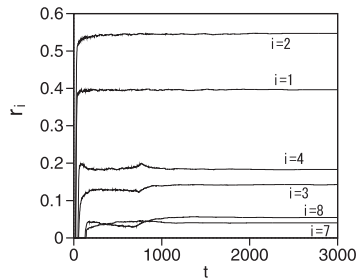}
\end{center}
\caption{Time evolution of the number ratio $r_i$ of $X_i$-dominant regions satisfying $X_i>1$ for $i=1,2,3,4,7$ and 8.
}
\label{f6}
\end{figure}
\begin{figure}[t]
\begin{center}
\includegraphics[width=12cm]{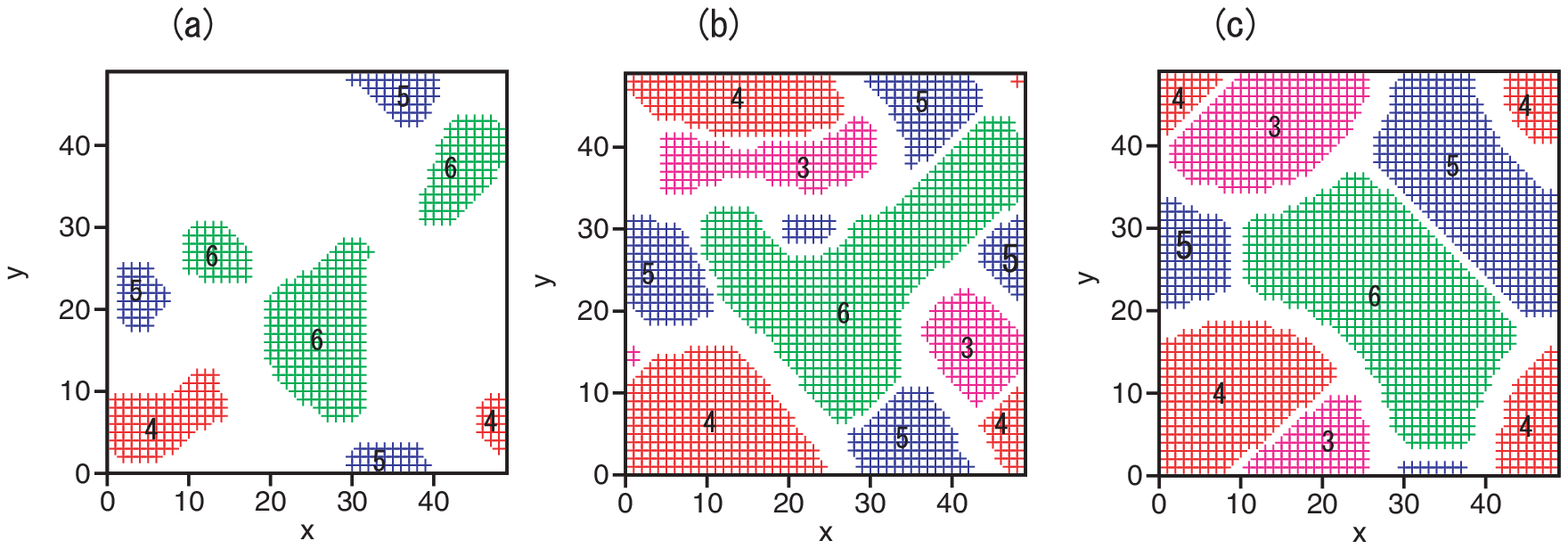}
\end{center}
\caption{Three snapshot patterns of the $X_i$ dominant region satisfying $X_i>1$ for $i=3.4.5$ and 6 at (a) $t=50$, (b) $t=100$, and (c) $t=2000$. The snapshot at $t=20000$ is shown in Fig.~5(b). The number $i$ indicates the cell type.
}
\label{f7}
\end{figure}

Finally, we have performed a numerical simulation in two dimensions.  The system size is $48\time 48$, and the periodic boundary conditions are imposed.  The network of the interaction is the same as before in Fig.~4(a), that is, the number of elements are fourteen and the layer number is three. The parameter values are $d_1=0.8$ and $d_2=1$ at $a=0.4,b=5,c=2,d_1=0.4,d_2=1,\alpha=0.4$ and $D_X=0.2$. 
The initial conditions are random between 0.5 and 0.6. 
Figure 5(a),(b) and (c) display patterns at $t=20000$ which represent the dominant region satisfying $X_i>1$ for each $i$ using different color for (a) $i=1$ and 2, (b) $i=3,\cdots,6$, and (c) $i=7,\cdots,14$. 
These states are final stationary states after the transient time evolution as shown later in Fig.~6 and Fig.~7. The whole region is divided into two domains: a $X_1$-dominant region and a $X_2$-dominant region. The $X_1$-dominant region is a circular blue region which is located near the four edges in Fig.~5(a). The area ratio of the $X_1$-dominant region and the $X_2$-dominant region is $0.42:0.58$, which is close to $d_1:d_2=0.44:0.56$. The $X_1$-dominant region is subdivided into two domains: a $X_3$-dominant region and a $X_4$-dominant region. The area ratio of the two regions is $0.44:0.56$, which is also close to the ratio $d_1:d_2=0.44:0.56$. Similarly, the $X_3$-dominant region is further subdivided into  a $X_7$-dominant region and a $X_8$-dominant region. The area ratio of the two regions is $0.42:0.58$, which is also close to the ratio $d_1:d_2=0.44:0.56$. Thus, the hierarchical differentiation is clearly seen. And it is shown that the area ratio is determined by the ratio of $d_1$ and $d_2$. 

Figure 6 displays the time evolution of the number ratio of the $X_i$-dominant regions for $i=1,2,3,4,7$ and 8. The number ratio becomes a stationary values at $t\sim 30$ for $i=1$ and 2, at $t\sim 70$ for $i=3$ and $4$, at $t\sim 1100$ for $i=7$ and 8. That is, the stationary state is attained sequentially from the upper layer to the lower layer.  Figures 7(a), (b) and (c) show three snapshot patterns of the $X_3$-,$X_4$-, $X_5$- and $X_6$-dominant regions at $t=50,100$ and $2000$. The dominant regions are initially small and random, and the size and the location are determined gradually. Figure 7(c) is almost similar to Fig.~5(b), which means that the time evolution is almost stationary at $t=2000$. 
\section{Summary}
Competitive interactions and hierarchical interactions are typical in gene networks.  We have proposed a simple cascade model of competitive reaction-diffusion equations for the cell differentiation. We have first reconfirmed a spatially-periodic pattern in a simple one-dimensional competitive reaction diffusion equations, which was originally proposed by Meinhardt and Gierer. It is a kind of the Turing pattern, but the origin of the formation of the periodic pattern is the long-range enhancement of the competitor. If the diffusion constant for $Y$-variable is assumed to infinity, we get a new model, in which a two-domain structure appears naturally from an random initial condition, and the size ratio of the two domains can be controlled by system parameters owing to the negative feedback effect of the domain size.  Next, we have generalized the model to a hierarchical model. The $X_i$-dominant regions appear in a cascade manner from the upper layer to the lower layer. The ratio of the $X_i$-dominant regions are well controlled by the system parameters. Our simple cascade model is not a realistic model based on biological experiments, however, it is a useful model to consider the Turing pattern, the ratio control, and the hierarchical differentiation in a unified manner.  We expect that our model might be applicable to specific cell differentiation processes by changing the reaction network and modifying the system parameters suitably.

\end{document}